\documentstyle[12pt]{article}
\newcommand{\beq}{\begin{eqnarray}}
\newcommand{\eeq}{  \end{eqnarray}}

\newcommand{\dv}[2]{\frac{\textstyle #1}{\textstyle #2}}

\def\prd#1 { Phys.\ Rev.\ D {\bf #1 }}
\def\npb#1 { Nucl.\ Phys.\ B {\bf #1 }}
\def\plb#1 { Phys.\ Lett.\ B {\bf #1 }}
\def\zpc#1 { Z.\@ Phys.\ C {\bf #1 }}
\def\prl#1 { Phys.\ Rev.\ Lett.\ {\bf#1 }}

\newcommand{\ra}{\rightarrow}

\newcommand{\vv}[1]{ {\bf #1}}
\def\lp{\lambda^{\prime} }

\def\an{ {J,J_3} }
\def\ji{ dx\ d^2{\bf k}_\bot}

\def\bsa{ $D^{\ast+}\ra D^0\pi^+$}
\def\bsb{ $D^{\ast+}\ra D^+\pi^0$}
\def\bsc{ $D^{\ast0}\ra D^0\pi^0$}
\def\bra{ $D^{\ast+}\ra D^+\gamma$}
\def\brb{ $D^{\ast0}\ra D^0\gamma$}
\def\brc{ $D^{\ast+}_{s}\ra D^{+}_{s}\gamma$}
\def\ta{ $D^{\ast +} \ra {\rm total}$ }
\def\tb{ $D^{\ast 0} \ra {\rm total}$ }
\def\Ba{ Br$^{a)}$}
\def\Bb{ Br$^{b)}$}
\def\Fa{ Eq.(\ref{e313})}
\def\Ga{ Eq.(\ref{e312a})}
\def\Fb{ Eq.(\ref{e313})}
\def\Gb{ Eq.(\ref{e312a})}

\def\GAa{ $\Gamma({\rm keV})^{a)}$ }
\def\GAb{ $\Gamma({\rm keV})^{b)}$ }

\begin{document}
\baselineskip=8mm
\newcount\sectionnumber
\sectionnumber=0

\begin{flushright}{ UTPT--94--15 }
\end{flushright}

\vspace{8mm}
\begin{center}
{\bf {\huge Strong and
Radiative  $D^\ast$  Decays } }\\
\vspace{6mm}
Patrick J. O'Donnell and Q. P. Xu\\
Physics Department,\\
University of Toronto,\\
Toronto, Ontario M5S 1A7, Canada.\\
\end{center}

\vskip 20pt
\centerline{\bf Abstract}

We use the relativistic light--front quark model to show that both
strong and radiative  $D^\ast$  decays are in good agreement with
the  1992  CLEO  II  results.  In  particular  the  coupling  for
$D^{\ast}\ra  D \pi$ is consistent  with the  experimental  upper
limit.  The key point is the relativistic  treatment of the quark
spin.

\newpage

\section{Introduction}

The $D^\ast$ decays has been under both
theoretical and experimental investigation
for quite a long time
\cite{pdg,accmor,cleo,kamal1,kamalx,cheng,cho,amudson,nardulli}.
The strong $D^\ast$ decays, $D^\ast \ra D \pi$, can be described
by the lowest--order (in external momentum) effective Lagangian
\beq
{\cal L} = (\dv{m_{D^\ast} g}{f_\pi})
{\hat D}^\dagger
\mbox{\boldmath $\tau$} \cdot\partial_\mu
\mbox{\boldmath $\pi$}  {\hat D}^{\ast\mu} \ ,
\label{en1}
\eeq
where ${\hat D}$ and ${\hat D}^\ast$ are the $D$ and $D^\ast$
isodoublet fields, and $\boldmath \tau$
is the isospin matrices.
The decay width of $D^{\ast +}\ra D^0 \pi^+$, for example, reads
\beq \Gamma_{D^{\ast +}\ra D^0\pi^+}
=\dv{g^2 |{\bf p_\pi}|^3}{6\pi f^2_\pi }\ .
\label{en4}
\eeq
The radiative $D^\ast$ decays $D^\ast \ra D \gamma$,
on the other hand, can be described by the decay amplitude
\beq
M(D^\ast \ra D \gamma)=
e \mu\ \epsilon_{\alpha\nu\sigma\lambda}\
\xi^{\ast\alpha} \epsilon^\nu p^\sigma_{D^\ast} p^\lambda_D \ ,
\label{b5}
\eeq
where $e \mu/2$ is the transition magnetic moment, and
$\xi^\alpha$ and $\epsilon^\nu$ are
the polarization vectors of the photon
and the $D^\ast$ meson, respectively. The radiative decay width
is
\beq
\Gamma=(e \mu)^2 \dv{|{\bf p_\gamma}|^3}{12\pi} \ .
\label{b7}
\eeq

In many theoretical studies  \cite{kamal1,kamalx}  people use the
non--relativistic  quark model to calculate the radiative $D^\ast$
decays  while  they  use  various  modifications  of the  $SU(4)$
symmetry  method for the strong $D^\ast$  decays.  Obviously, the
treatment of the strong and the radiative  $D^\ast$ decays is not
on the same footing.

However one can also use the quark model to calculate  the strong
decays.  Through  the PCAC  relation  one can  relate  the strong
coupling $g$ to the axial--current form factor $A_0(0)$
\beq
g=A^{D D^\ast}_0(0)\ .
\label{en8}
\eeq
Here $A_0$ is one of the four form  factors  used to describe the
matrix element of a vector meson $V$ decaying to a  pseudo--scalar
meson $X$ by the quark transition $Q^\prime \ra Q$:
\begin{eqnarray}
&&\hspace{-1cm}\langle
 X(p_X) \!\!\mid {\bar Q} \gamma_\mu (1-\gamma_5) Q^\prime
\mid \!\! V(p_V,\epsilon) \rangle
=\!\!\dv{2 V(q^2)}{m_X+m_V} i \epsilon_{\mu\nu\alpha\beta}
\epsilon^{\ast\nu} p_X^\alpha p_V^\beta -\!\! 2m_V
\frac{(\epsilon^\ast \cdot p_X) }{q^2} q^\mu A_0(q^2)\nonumber\\
&&\hspace{-1cm}-\!\left[ (m_X+m_V) \epsilon^{\ast\mu} A_1(q^2)\!-
\frac{(\epsilon^\ast \cdot p_X) }{m_X+m_V}(p_X+p_V)^\mu A_2(q^2)\!
-2m_V \frac{(\epsilon^\ast \cdot p_X)}{q^2} q^\mu A_3(q^2)\right] ,
\label{eb2}
\end{eqnarray}
where $\epsilon$ is
the polarization vector of the vector meson, $q=p_V-p_X$, and
\beq
A_3(q^2)=\dv{m_X+m_V}{2 m_V} A_1(q^2)-\dv{m_V-m_X}{2 m_V} A_2(q^2)
\ , \ A_3(0)=A_0(0)\ .
\label{eb3}
\eeq
In Eq.\ (\ref{en8}), $A^{D D^\ast}_0(0)$ is the $A_0(q^2=0)$ form
factor for $D^\ast \ra D$ induced by the quark  transition $u \ra
d$.  In the  non--relativistic  quark model, one obtains $g \simeq
1$  \cite{yan}.  In a  rather  different  model,  the  BSW  model
\cite{bsw}, one has \cite{kamalx}
\beq
g=A^{D D^\ast}_0(0)=\int dx\ d^2{\bf k}_\bot
\phi^{\ast}_{D}(x,\vv{k_\bot})
\phi_{D^\ast}(x,\vv{k_\bot})\,\,\,  ,
\label{g1}
\eeq
where  $\phi(x,\vv{k_\bot})$ is the momentum wavefunction.  Since
the wavefunctions of the $D$ and $D^\ast$ are quite similar, this
model  also has $g  \simeq  1$.  In fact,  in the  heavy  $c$--quark
limit             $\phi_{D^\ast}(x,\vv{k_\bot})$             and
$\phi_{D}(x,\vv{k_\bot})$  should be exactly  the same, due to the
heavy quark  symmetry,  and hence in the BSW model, $g=1$ in this
limit.  These results of $g$ are, however,  larger than the upper
limit   of   $g$   obtained    from   the   ACCMOR    measurement
$\Gamma_{D^{\ast +}}<131$keV \cite{accmor}:
\beq
g < 0.7 \ .
\label{en0}
\eeq

There  have been  some  recent  studies  on the  $D^\ast$  decays
\cite{cheng,cho,amudson,nardulli}.  In   ref.  \cite{cheng}   the
radiative $D^\ast$ decays are treated with the  non--relativistic
quark  model  but  the  strong  coupling  $g$ is  related  to the
quark--level  axial--current  form  factor  $g^{ud}_A=\dv{3}{4}$;
this is obtained from fitting the measured nucleon axial--current
form factor  $g^{\rm  nucleon}_A=1.25=\dv{5}{3}  g^{ud}_A$.  Thus
the   coupling   $g$   is    $g=g^{ud}_A=\dv{3}{4}$.   In   refs.
\cite{cho,amudson}   heavy--meson   chiral--perturbation   theory
\cite{chiral,yan}  is used  and all the  $D^\ast$  decays  can be
described by two parameters.  These  parameters are the couplings
of the chiral  Lagrangian and can therefore be determined only by
fitting   the    experimental    data.   In   a   recent    study
\cite{nardulli},  the radiative  decays are calculated  using the
vector--dominance  hypothesis  while the strong  coupling  $g$ is
extracted  from the data on the decay  $D\ra \pi e {\bar  \nu}_e$
using heavy--meson chiral--symmetry relations.  The central value
obtained  for  $g$  is  $g  \approx   0.4$.  In  this  study  the
uncertainties  associated  with  assumptions  on the  $f_+$  form
factor  of $D\ra \pi e {\bar  \nu}_e$  and on the decay  constant
$f_D$,  etc.,  could be  large.  A value of $g \sim  0.4$ is also
obtained  \cite{bob}  in a numerical  solution to another type of
relativistic quark model \cite{bs}.

The  failure  of the  non--relativistic  quark  model and the BSW
model   in   explaining    $g<0.7$   has   its   roots   in   the
non--relativistic  treatment  of the quark spin in these  models.
In this  letter,  we  present a  calculation  of the  strong  and
radiative decays of $D^\ast$ using the relativistic  light--front
quark model  \cite{Russian,jaus,ox}.  In our  calculation  we are
able to treat on the same basis  both the strong and  radiative
decays.  Due to the relativistic  treatment one
can  see  unambiguously  how  the  strong  coupling  $g$  can  be
consistent  with the limit in Eq.\  (\ref{en0}).  Our  calculated
branching  ratios for the strong and radiative decays are in good
agreement with the 1992 CLEO II measurement \cite{cleo}.

\section{The light--front relativistic quark model}

The  relativistic  light--front  quark model
was  developed  quite a long  time ago and  there  have been many
applications   \cite{Russian,jaus,ox}.  Here  we   give  a  brief
introduction.

A  ground--state  meson $V({\bar Q}q)$ with spin $J$ on the light
front can be described by the state vector
\beq
| V (P, J_3, J) \rangle=\!\!\!\!\!&&\!\!\!\!\!\int d^3\vv{p_1} d^3\vv{p_2}
\ \delta(\vv{P}-\vv{p_1}-\vv{p_2}) \nonumber\\
\!\!\!\!&&\!\!\sum_{\lambda_1,\lambda_2}
\Psi^\an(\vv{P},\vv{p_1},\vv{p_2},\lambda_1,\lambda_2)
|Q(\lambda_1,\vv{p_1} )\ {\bar q}(\lambda_2,\vv{p_2} )\rangle\ ,
\label{e31}
\eeq
In the light--front convention, the quark coordinates are
given by
\beq
&&p_{1}^+=x_1 P^+ \ , p_{2}^+=x_2 P^+ \ ,
\ x_1+x_2=1 \ , 0\le x_{1,2}\le 1 \ ,\nonumber\\
&&\vv{p_{1\bot}}=x_1 \vv{P_\bot}+\vv{k_\bot} \ , \
\vv{p_{2\bot}}=x_2 \vv{P_\bot}-\vv{k_\bot} \ .
\label{e32}
\eeq
The quantities $x_{1,2}$ and ${\bf k}^2_\bot$ are invariant
under the kinematic Lorentz transformations.
Rotational invariance of the wave function for states with spin
$J$ and zero orbital angular momentum requires the wave function
to have the form \cite{Russian,jaus} (with $x=x_1$)
\beq
\Psi^{J,J_3}(\vv{P},\vv{p_1},\vv{p_2},\lambda_1,\lambda_2)
=R^\an(\vv{k_\bot},\lambda_1,\lambda_2)
\phi(x, \vv{k_\bot}),
\label{e36}
\eeq
where $\phi(x, \vv{k_\bot})$ is even in $\vv{k_\bot}$ and
\beq
R^{J,J_3}(\vv{k_\bot},\lambda_1,\lambda_2)=\!\!
\sum_{\lambda,\lp}
\langle \lambda_1 | R^\dagger_M(\vv{ k_\bot},m_Q) |\lambda \rangle
\langle \lambda_2 | R^\dagger_M(\vv{-k_\bot},m_{\bar q}) |\lp \rangle
C^{J,J_3}(\dv{1}{2},\lambda;\dv{1}{2},\lambda^\prime).
\label{e37}
\eeq
In Eq.\ (\ref{e37}),
$C^{J,J_3}(\dv{1}{2},\lambda;\dv{1}{2},\lambda^\prime)$ is
the Clebsh--Gordan coefficient and the rotation
$R_M(\vv{k_\bot}, m_i)$
on the quark spins is the Melosh rotation \cite{melosh}:
\beq
R_M(\vv{k_\bot},m_i)=
\dv{m_i+x_i M_0-i
\mbox{\boldmath $\sigma$ }\vv{\cdot}(\vv{n\times k_\bot})}
{\sqrt{ (m_i+x_i M_0)^2+\vv{k}^2_{\bot} } } \ ,
\label{e33}
\eeq
where $\vv{n}=(0,0,1)$, $\boldmath\sigma$ is the Pauli spin matrix,
and
\beq
M_0^2=\dv{m_Q^2+\vv{k}^2_{\bot}}{x_1}+\dv{m_{\bar q}^2+\vv{k}^2_{\bot}}{x_2} \
{}.
\label{e34}
\eeq
The spin wave function
$R^{J,J_3}(\vv{k_\bot},\lambda_1,\lambda_2)$ in Eq.\ (\ref{e37}) can also
be written as
\beq
R^{J,J_3}(\vv{k_\bot},\lambda_1,\lambda_2)
\!\!\!\!\!&&=
\chi^\dagger_{\lambda_1} R^\dagger_M(\vv{ k_\bot},m_Q)
\ S^\an\ R^{\dagger T}_M(\vv{-k_\bot},m_{\bar q}) \chi_{\lambda_2}
\nonumber\\
\!\!\!\!\!&&=\chi^\dagger_{\lambda_1}\ U^\an_V\ \chi_{\lambda_2} \ ,
\label{e35a}
\eeq
where $S^\an$ is defined by
\beq
S^{J,J_3}=\sum_{\lambda,\lambda^\prime}
|\lambda\rangle \langle \lambda^\prime | \
C^{J,J_3}(\dv{1}{2},\lambda;\dv{1}{2},\lambda^\prime) \ .
\label{e35b}
\eeq
For the pseudoscalar and vector mesons,
\beq
S^{0,0}=\dv{i \sigma_2}{\sqrt2} \ \ , \ \
S^{1,\pm 1}=\dv{1\pm\sigma_3}{2}\ \ , \ \
S^{1,0}=\dv{\sigma_1}{\sqrt2} \ .
\label{e27}
\eeq
The explicit expressions of $U^{1,J_3}$ in Eq.\ (\ref{e35a}) can
be found in ref. \cite{ox}.

The matrix element of a vector  meson $V(Q^\prime {\bar q})$
decaying to a pseudo--scalar  meson $X(Q{\bar q})$ is
\beq
\!\!\!\!\langle X(p_X) |\bar{Q}
\Gamma Q^\prime |  V(p_V,J_3)\rangle \!\!\!\!
&&\!\!\!\!=\int dx\ d^2\vv{k_\bot}
\sum_{\lambda_1,\lambda_2}
\dv{ \Psi^{\ast 0,0}_X\ \bar u_Q \Gamma u_{Q^\prime}\ \Psi^{1,J_3}_V}
{x}
\nonumber\\
\!\!\!\!&&\!\!\!\!=\int dx\ d^2\vv{k_\bot}
\dv{ \phi^\ast_X \phi_V }{x}
Tr\left[ U^{\dagger 0,0}_X\ U_\Gamma\ U^{1,J_3}_V \right]\ ,
\label{e38}
\eeq
where $U_\Gamma$ is defined by
\beq
{\bar u}^i_Q\ \Gamma\ u^j_{Q^\prime}=\chi^\dagger_i\ U_\Gamma\ \chi_j \ ,
\label{e39}
\eeq
In Eq.\ (\ref{e38}), we choose
$q^+=0$ so $q^2=-{\bf q}^2_\bot$.
In contrast to Eq.\ (\ref{e38}),
the matrix element in the BSW model is given by
\beq
\langle  X(p_X) |\bar{Q}
\Gamma Q^\prime | V(p_V,J_3) \rangle
\!\!=\int \ji
\dv{\phi^\ast_X \phi_V}{x}
Tr\left[ S^{\dagger 0,0} U_\Gamma S^{1,J_3} \right]   \ .
\label{e310}
\eeq

Eq.\ (\ref{e38}) is in fact expected to be valid only for ``good"
currents  such as  $\Gamma=\gamma^+,\gamma^+\gamma_5,  $  \ldots.
There  are  contributions  other  than  the  one  given  in  Eq.\
(\ref{e38}) if a current is not a ``good" current \cite{jaus}.

Using the ``good" currents
$\Gamma=\gamma^+\gamma_5, \gamma^+$ and Eq.\ (\ref{e38}),
we obtain the following expressions for the
form factors $A_0(0)$ and $V(0)$ of the $Q^\prime \ra Q$ transition
\beq
\hspace{-0.55cm}A_0(0)\!\!\!\!&=&\!\!\!\!\!\!
\int\!\!\dv{ dx\ d^2\vv{k_\bot} \phi^\ast_X\phi_V}
{\sqrt{(A^2_V\!+\!{\bf k}^2_\bot)(A^2_X\!+\!{\bf k}^2_\bot)}}
\!\left[ A_V A_X\!+\!(2x-1){\bf k}^2_\bot+\!
\dv{2 (m_{Q^\prime}+m_Q) (1-x){\bf k}^2_\bot}{W_V} \right]
\label{a1}\\
\hspace{-0.55cm}V(0)\!\!\!\!&=&\!\!\!\!\!\!
\int\!\!\dv{dx\ d^2{\bf k}_\bot \phi^\ast_X\phi_V}
{\sqrt{(A^2_V\!+\!{\bf k}^2_\bot)(A^2_X\!+\!{\bf k}^2_\bot)}}
(m_X\!+\!m_V)(1\!-\!x) \nonumber\\
&&\phantom{abcdefghijklmnabcd}
\left[ A_X\!+\!\dv{{\bf k}^2_\bot}{W_V}\!
+(1-x)(m_Q-m_{Q^\prime}) {\bf k}^2_\bot \theta_V \right]\
\label{a2}
\eeq
where
\beq
&&A_V=x\ m_{\bar q}+(1-x) m_{Q^\prime}\ , \ A_X=x\ m_{\bar q}+(1-x) m_Q
\nonumber\\
&&W_V=M^V_0+m_Q+m_{\bar q}\ , \
\theta_V=(\dv{d\phi_V}{d{\bf k}^2_\bot} )/\phi_V \ .
\label{a3}
\eeq
Here, $M^V_0$ corresponds to Eq.\ (\ref{e34}) for the meson $V$.

The  meson  wave   functions   $\phi(x,\vv{k_\bot})$   are  model
dependent and difficult to obtain; often simple forms are assumed
for them.
One possibility is to use the wavefunction adopted in \cite{bsw}
\beq
\phi(x,\vv{k_\bot})=
N \sqrt{x(1-x)}
exp\left( -\ \dv{M^2}{2w^2}
\left[ x-\dv{1}{2}-\dv{m_Q^2-m_{\bar q}^2}{2 M^2}\right]^2\right)
\dv{ exp\left(-\ \dv{ \vv{k}^2_{\bot} }{2w^2}\right)}
{\sqrt{\pi w^2} } \ ,
\label{e313}
\eeq
where $M$ is the mass of the meson.  In  \cite{jaus} a Gaussian  type of
wavefunction  was
used,
\beq
\phi(x,\vv{k_\bot})=N\sqrt{\dv{d\vv{k}_z}{dx} }exp\left(-\dv{\vv{k}^2}{2
\omega^2} \right) \ ,
\label{e312a}
\eeq
where $\vv{k}_z$ is defined by $x_1 M_0=E_Q+k_z$ with
$E_Q=\sqrt{ \vv{k}^2_{\bot}+k^2_z+m_Q^2}$. The  parameter  $\omega$ in
both Eqs.\  (\ref{e313}) and (\ref{e312a}) should be of the
order of  $\Lambda_{\rm  QCD}$.  As we will see, the  results are
not too sensitive to the choice of wavefunction.

It is  interesting  to  look  at  the  general  behavior  of  the
wavefunction  $\phi(x,\vv{k_\bot})$  in the  limit  of  $m_Q  \ra
\infty$.  The   distribution   amplitude   $\int   d^2\vv{k}_\bot
\phi(x,\vv{k}_\bot)$  of a heavy meson, as is well known,  should
have a peak near $x\simeq  x_0=\dv{m_Q}{m_V}$.  As $m_Q$  becomes
larger the width of the peak  decreases and $x_0$ comes closer to
1.   The   wavefunction    $\phi(x,\vv{k}_\bot)$    vanishes   if
$\vv{k}^2_\bot  >>  \Lambda^2_{\rm  QCD}$  and in  the  $m_Q  \ra
\infty$   limit   peaks   at   $x_0\ra   1$  as  the   width   of
$\phi(x,\vv{k}_\bot) \ra 0$.  Both wavefunctions listed have this
feature.

\section{The $D^\ast$ decays}

We now  calculate  the $D^\ast$  decay rates in the  light--front
quark  model.  For $D^\ast  \ra D \pi$, we first look at the form
factor  $A_0(0)$ of the  transition $Q \ra Q$.  In this case Eq.\
(\ref{a1}) reduces to
\beq
A^{ {\bar Q}Q}_0(0)=
\int dx\ d^2\vv{k_\bot} \phi^\ast_X\phi_V T_{A_0}
\label{b1}
\eeq
where
\beq
T_{A_0}=\dv{ A^2+(2x-1){\bf k}^2_\bot+
\dv{4 m_Q (1-x)\vv{k^2_\bot}}{W_V} }
{A^2+{\bf k}^2_\bot}
\label{b2}
\eeq
and
\beq
A=x\ m_{\bar q}+(1-x)\ m_Q \ .
\label{b2b}
\eeq

Generally  $T_{A_0} < 1$ since it comes  directly from the Melosh
rotation.  (For no Melosh  rotation  one would get  $T_{A_0}=1$.)
How far $T_{A_0}$  deviates from 1 depends on the relation  among
$m_Q$, $m_{\bar q}$ and ${\bf k}^2_\bot$, where the average value
of ${\bf  k}^2_\bot$ is determined by the scale  parameter in the
wavefunction  such as the  $\omega$  in  Eqs.\  (\ref{e313})  and
(\ref{e312a}).  Usually as $< {\bf  k}^2_\bot>$  becomes  smaller
$T_{A_0}$  becomes  closer  to 1.  If  ${\bf  k}_\bot=0$  in Eq.\
(\ref{b2}) then $T_{A_0}\equiv 1$.

Let us consider two limiting cases for the transition  quark $Q$.
If  the  transition  quark  $Q$ is  infinitely  heavy  while  the
spectator quark is light, $x \ra 1$, and therefore the third term
in the numerator of $T_{A_0}$ vanishes.  Thus
\beq
T_{A_0} \ra 1 \ \ {\rm and} \ \
A^{ {\bar Q}Q}_0(0) \ra \int\ji\phi^{\ast}_{X}\phi_{V}
\ra 1 \ ,
\label{b3}
\eeq
as  required  by  the  heavy  quark  limit.  Note  in  this  case
$q^2=q^2_{max}=0$.  If, as in the decays  $D^\ast \ra D \pi$, the
transition quark $Q$ is light while the spectator quark is heavy,
then in Eqs.\  (\ref{b1})  and  (\ref{b2}), $x \ra 0$ as $m_c \ra
\infty$.  In this case the third term in the  numerator  of  $T_{A_0}$
also vanishes, but now
\beq
T_{A_0} \ra \dv{A^2-{\bf k}^2_\bot}{A^2+{\bf k}^2_\bot}\ .
\label{b4}
\eeq
Since both ${\bf  k}^2_\bot$ and $A^2$ are of the same order, one
can expect  considerable  deviation  from 1 for  $A_0(0)$ in this
case.

Obviously, the situation for $g=A^{D D^\ast}_0(0)$ is quite close
to the second case  discussed  above.  Thus, one can clearly  see
why the light--front model gives a deviation of $g$ from 1.

To calculate the radiative decays $D^\ast \ra D \gamma$,
one can express $\mu$ in Eq.\ (\ref{b5}) in terms of
the form factor $V(0)$
\beq
\mu=[ e_c V^{{\bar c}c}(0)+e_q V^{{\bar q}q}(0)]
(\dv{2}{m_D+m_{D^\ast}})
\label{b8a}
\eeq
where  $q=u$\  or\  $d$  and  $e_c=\dv{2}{3}$,   $e_u=\dv{2}{3}$,
$e_d=-\dv{1}{3}$.  $V^{ {\bar  Q}Q}(0)$ is the $V(0)$ form factor
of the  transition  $Q\ra Q$ ($Q=c, u\ {\rm or}\ d$).  $V^{ {\bar
Q}Q}(0)$  in the  light--front  model  can be obtained from  Eq.\
(\ref{a2})
\beq
V^{ {\bar Q}Q}(0)= (m_D+m_{D^\ast})
\int \dv{dx\ d^2\vv{k_\bot}\phi^\ast_D \phi_{D^\ast} }
{A^2+{\bf k}^2_\bot}
(1-x)\left\{ A+\dv{{\bf k}^2_\bot}{W_{D^\ast}} \right\}\ .
\label{b9}
\eeq

To calculate $V^{{\bar c}c}(0)$,  $V^{{\bar q}q}(0)$, and $g$, we
use  the  two  wavefunctions  given  in  Eqs.\  (\ref{e313})  and
(\ref{e312a})  as the  wavefunctions  for  the  $D$ and  $D^\ast$
mesons.  We assume  $m_u=m_d=m$ and expect that the  light--quark
masses $m$ in the  wavefunctions to be around $0.25\sim  0.30{\rm
GeV}$.  Similarly   the   scale   parameter   $\omega$   in  both
wavefunctions  should be around $0.40\sim  0.50{\rm GeV}$.  These
values are similar to those taken in \cite{bsw} and  \cite{jaus}.

The  dependence of $V^{ {\bar Q}Q}(0)$ on the quark mass $m_Q$ is
similar    to    that    in    the    non--relativistic     quark
model \cite{ISGW,OT}:
$V^{   {\bar    Q}Q}=(\dv{m_D+m_{D^\ast}}{2} )  \sqrt{
\dv{m_{D^\ast}}{m_D}  } \ ( \dv{1}{m_Q} )\ $.  That is $V^{ {\bar
Q}Q}(0)$  decreases as $m_Q$ increases and if $m_Q \ra \infty$, $V^{
{\bar Q}Q}(0) \ra 1$, as required by the heavy quark relation.

In the figure we show the  dependence  of the coupling $g$ on the
light  quark mass $m$ and  $\omega$  in the  wavefunctions  Eqs.\
(\ref{e313})  and  (\ref{e312a}).  We  choose  $m_c=1.6$GeV;  the
coupling  $g$ has little  dependence  on $m_c$.  One can see that
for a fixed  $\omega$,  g  increases  as $m$  increases.  For $m$
between  $0.25 \sim  0.30$GeV  and  $\omega$  between  $0.40 \sim
0.45$GeV,  $g$ is between  $0.55 \sim 0.65$.  To avoid  adjusting
parameters, we choose  $m=0.25$GeV, and  $\omega=0.4$GeV  for the
calculation of decay rates.  There is good agreement  between the
calculated  branching  ratios  and  experiment.  With this set of
parameters $g=0.6$ using Eq.\ (\ref{e313}) or Eq.\ (\ref{e312a}).
In the table , we show the corresponding  results for $D^\ast \ra
D \pi$ and $D^\ast \ra D \gamma$.  We also show how the branching
ratios change with a different  choice of mass.  Finally, we give
the  results  of  our  calculation  for  the  decay  \brc  \ (for
$m_s=0.40$GeV and  $m_s=0.50$GeV).  This decay is usually assumed
to be the dominant mode.  In a recent paper \cite{cw} it has been
argued that through isospin symmetry breaking there can also be a
significant    correlation    between   the   branching    ratios
Br($D^{\ast}_s  \ra D_s \pi^0$) and Br($D^{\ast +} \ra D^+ \gamma
$).

\section{Conclusion}

In this letter we have presented a calculation of both strong and
radiative  $D^\ast$  decays  using the  relativistic  light--front
quark  model.  Our results  are in good  agreement  with the 1992
CLEO II  measurement.  In particular  the strong  coupling $g$ is
consistent  with the  experimental  upper limit  $\Gamma_{D^{\ast
+}}<131$keV.  The key point is the relativistic  treatment of the
quark  spin.  The  fact  that  the   relativistic   treatment  is
essential  in  explaining  $g<1$ is  reminiscent  of the  similar
situation  in  the  nucleon  axial--current  form  factor  $g^{\rm
nucleon}_A$ where the non--relativistic  quark model gives $g^{\rm
nucleon}_A=\dv{5}{3}$   and  a  relativistic  treatment  can  get
$g^{\rm   nucleon}_A$   naturally  down  to  the  measured  value
\cite{close}.

\vspace{.3in}
\centerline{ {\bf Acknowledgment}}
We thank Felix Schlumpf for useful correspondence.
This work was in part supported by the Natural Sciences and Engineering
Council of Canada.

\newpage

{\bf Figure:}

The dependence of the coupling $g$ on the  light--quark  mass $m$
and  $\omega$  in Eqs.\  (\ref{e313})  and \  (\ref{e312a}).  The
solid lines are the results of Eq.  (\ref{e312a})  and the dashed
lines use Eq.  (\ref{e313}).  The upper  solid and  dashed  lines
correspond to $\omega = 0.40$ GeV, and the lower solid and dashed
lines are for $\omega = 0.45$ GeV.

\newpage

\baselineskip=25pt

\begin{table}
\baselineskip=25pt

\vspace{1.5cm}
\begin{center} Table.
Numerical results for $D^\ast$ decays.
Also listed are the experimental results of
CLEO II \cite{cleo} and PDG \cite{pdg}. All branching ratios
are in $\%$.
\end{center}

\begin{center}
\begin{tabular}{ l c c  c c c c } \hline\hline
 Decay& \Ba & \Ba & \Bb & \Bb &        Br             &    Br     \\
      &\Fa  & \Ga & \Fb &\Gb  &      CLEO II          &   PDG     \\
 \hline
 \bsa & 68.2& 68.0& 68.1& 67.7&68.1$\pm$ 1.0$\pm$ 1.3 & 55$\pm$ 4 \\
 \bsb & 31.6& 31.5& 31.6& 31.4&30.8$\pm$ 0.4$\pm$ 0.8 & 27.2$\pm$ 2.5\\
 \bra &  0.3&  0.5&  0.4&  0.9& 1.1$\pm$ 1.4$\pm$ 1.6 & 18$\pm$ 4 \\
\hline
 \bsc & 75.2& 73.0& 70.5& 66.2&63.6$\pm$ 2.3$\pm$ 3.3& 55$\pm$ 6  \\
 \brb & 24.8& 27.1& 29.5& 33.8&36.4$\pm$ 2.3$\pm$ 3.3& 45$\pm$ 6  \\
\hline\hline
      &\GAa &\GAa &\GAb &\GAb &                      &            \\
      &\Fa  &\Ga  &\Fb  &\Gb  &                      &            \\
\hline
\ta   &122.8&123.0&104.9&102.1&                      &            \\
\tb   & 74.0& 76.3& 67.4& 69.5&                      &            \\
\hline
\brc  &  0.1&  0.1&  0.2&  0.3&                      &            \\
\hline\hline
\end{tabular}
\end{center}

\vskip 1cm

$a)$: $\omega=0.4$GeV, $m=m_u=m_d=0.30$GeV, $m_s=0.50$GeV.\\
$b)$: $\omega=0.4$GeV, $m=m_u=m_d=0.25$GeV, $m_s=0.40$GeV.\\

\end{table}

\end{document}